# Relative Time-of-Flight Measurement in an Ultrafast Electron Microscope

Jialiang Chen[1,2,†], Simon A. Willis[1,2], and David J. Flannigan[1,2,*]

[1]*Department of Chemical Engineering and Materials Science, University of Minnesota, 421 Washington Avenue SE, Minneapolis, MN 55455, USA*

[2]*Minnesota Institute for Ultrafast Science, University of Minnesota, Minneapolis, MN 55455, USA*

**Abstract:** Efforts to push the spatiotemporal imaging-resolution limits of femtosecond (fs) laser-driven ultrafast electron microscopes (UEMs) to the combined angstrom-fs range will benefit from stable sources capable of generating high bunch charges. Recent demonstration of unconventional off-axis photoemitting geometries are promising, but connections to the observed onset of structural dynamics are yet to be established. Here we use the *in-situ* photoexcitation of coherent phonons to quantify the relative time-of-flight (*r*-TOF) of photoelectron packets generated from the Ni Wehnelt aperture and from a Ta cathode set-back from the aperture plane. We further support the UEM experiments with particle-tracing simulations of the precise electron-gun architecture and photoemitting geometries. In this way, we measure discernable shifts in electron-packet TOF of tens of picoseconds for the two photoemitting surfaces. These shifts arise from the impact the Wehnelt-aperture off-axis orientation has on the electron-momentum distribution, which modifies both the collection efficiency and the temporal-packet distribution relative to on-axis emission. Future needs are identified; we expect this and other developments in UEM electron-gun configuration to expand the range of materials phenomena that can be directly imaged on scales commensurate with fundamental structural dynamics.

[*]Author to whom correspondence should be addressed.
Email:  flan0076@umn.edu
Office:  +1 (612) 625-3867

[†]Currently at Analog Devices, Inc., San Jose, CA

Femtosecond (fs) laser-based ultrafast electron microscopes (UEMs) employing (Schottky) field-emission or thermionic electron guns (S/FEGs or TEGs, respectively) can be operated via photoelectron emission from a photocathode source.[1–14] Generally, a visible or UV pulse train is focused onto a photoemitting material located at the position of the conventional electron source. For UEMs employing Wehnelt electrodes, the photoemitting material is typically set-back from the plane of the Wehnelt aperture.[11,15–17] Particularly when using UV photons, or when the laser-spot size is larger than the cathode tip apex, electrons can be photoemitted from multiple surfaces at different planar distances from the anode plane.[11,12,18–20] This can lead to the generation of more than one discrete electron packet from a single UV fs laser pulse and, thus, a difference in arrival time at the specimen.[11,19] Scattering information gathered from such a condition will produce data points that are convolutions of more than one discrete time point if the packet distributions are well-resolved in time. If they are not resolved, it will still lead to an effectively broadened packet and degradation of temporal resolution.[11,12,21]

Accordingly, photoemission from a single point, as for S/FEGs, or from a single planar surface, as for TEGs, is desirable in UEM. This is not to say that selectable, spatially-confined emission from different regions in the gun cannot be useful.[12] For example, we have found that photoemission from the Wehnelt aperture surface can produce stable and robust pulsed electron beams of comparable quality to that from on-axis photocathodes.[18] Importantly, however, such geometries beg the question of time-zero shift relative to photoemission from the flat photocathode position. A few methods for determining electron-packet arrival time at the specimen plane in fs UEM have been demonstrated.[22–27] Of these, the photon-induced near-field effect has been used to illustrate relative shifts in packet arrival time when photoemitting from different parts of the cathode in a TEG-based UEM with independent Wehnelt biasing.[11]

Here, we use UEM imaging of photoexcited coherent acoustic phonons (CAPs) to quantify the relative time-of-flight ($r$-TOF) of photoelectron packets generated from the anode-facing surface of a Ni Wehnelt aperture and from the flat tip apex of a Ta cathode set back from the aperture plane. For laser-pulse trains aligned entirely on the photoemitting surfaces, we measure a relative difference in TOF of 37.3 ± 0.5 ps for a 0.35-mm set-back position. This manifests in the initial photoexcitation of coherent $c$-axis phonons in a multilayer flake of 2$H$-MoS$_2$. That is, CAP dynamics probed with photoemission from the Wehnelt aperture are shifted forward in time due to the shorter pathlength of the UV pulse train relative to the Ta cathode. General Particle Tracer (GPT) simulations of the specific experimental geometry, in tandem with field maps calculated using Poisson Superfish, mostly agree with the experiments with some notable deviations.[28,29]

Figure 1 shows the photoemitting geometries and the specimen used to quantify $r$-TOF with CAP dynamics. The 2$H$-MoS$_2$ specimen was prepared following previously reported methods.[30,31] The cathode was a truncated, 0.2-mm diameter polycrystalline Ta source (Applied Physics Technologies). For all measurements, the source was set-back 0.35 mm from a 1.0-mm diameter Ni Wehnelt aperture (Thermo Fisher, Fig. 1a). Photoelectrons from both surfaces were generated with 240-fs pulses (fwhm) of 4.8-eV photons, a repetition rate ($f_{rep}$) of 20 kHz, and a pulse energy of 75 nJ (Light Conversion PHAROS and HIRO). The UV laser-spot size on the photoemitting surfaces was estimated to be 50 µm (fwhm).[32] The position of the UV laser-pulse train on the emitting surfaces was controlled with a piezo mirror positioner in the gun periscope module of the Tecnai Femto UEM (Thermo Fisher). Gun alignments were separately optimized for the two photoemitting geometries. Images were acquired with a 30-s acquisition time using a Gatan OneView 16 MP CMOS camera.[33] For all scans, the specimen was photoexcited with 2.4-eV photons, an $f_{rep}$ of 20 kHz, and a fluence of 10.8 mJ/cm$^2$.

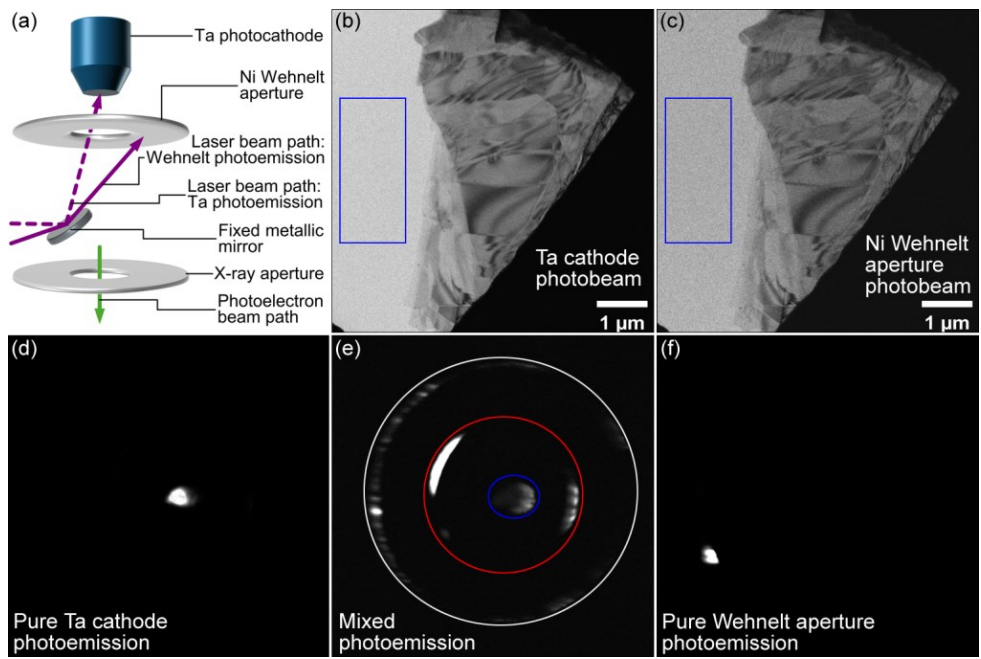

**Figure 1.** The two photoemission geometries and the 2$H$-MoS$_2$ specimen used in the electron-packet $r$-TOF measurements. (a) Simplified schematic of key elements of the UEM electron gun and the two photoemission geometries (anode and dynodes omitted for clarity). (b,c) Pulsed-beam bright-field image of the 2$H$-MoS$_2$ flake obtained with the Ta cathode photobeam and with the Ni Wehnelt aperture photobeam, respectively. Blue rectangles outline the regions from which image statistics were generated (see main text). (d) Image of pure Ta cathode photoemission. (e) Image of photoemission from both the Ta cathode (blue = tip, red = shank) and the Wehnelt aperture (white). (f) Image of pure Wehnelt-aperture photoemission.

Prior to conducting UEM imaging scans with the photoemitting surfaces, we verified that pure photoemission from those surfaces could be achieved and could produce viable and comparable images. While the signal-to-noise ratio (*SNR*) was lower for aperture photoemission (with the same UV pulse energy), the bright-field contrast patterns were similar (Fig. 1b,c), and the qualities were comparable and usable for quantitative analysis of CAP dynamics from the same regions of interest. Image *SNR* was calculated using Equation 1.

$$SNR = \left\{ \sum_{i=1}^{N} \sum_{j=1}^{M} \hat{I}(i,j)^2 / \sum_{i=1}^{N} \sum_{j=1}^{M} [I(i,j) - \hat{I}(i,j)]^2 \right\}^{1/2} \quad (1)$$

Here, $I$ is the input $M \times N$ image size, and $\hat{I}$ is the noise-free image of $I$. We estimated $\hat{I}$ from the vacuum regions of approximately constant intensity outlined in blue in Figure 1b,c by assuming entirely Gaussian image noise. We then assumed $\hat{I}$ to be equal to the average intensity, $\bar{I}$, within the associated outlined regions, as per Equation 2.

$$\hat{I}(i,j) \cong \bar{I} = \frac{1}{MN} \sum_{i=1}^{N} \sum_{j=1}^{M} I(i,j) \quad (2)$$

Image standard deviation, $\sigma(I)$, was determined from the outlined regions using Equation 3.

$$\sigma(I) = \left\{ \frac{1}{MN} \sum_{i=1}^{N} \sum_{j=1}^{M} [I(i,j) - \bar{I}]^2 \right\}^{1/2} \quad (3)$$

We estimated *SNR* from Equations 2 and 3 $\left( SNR \cong \bar{I}/\sigma(I) \right)$. Note that the lower *SNR* for aperture photoemission is likely due to a combination of lower quantum efficiency and the off-axis geometry. (The lower collection efficiency at the X-ray aperture as a result of the off-axis geometry also appears in the GPT simulations.[16,17])

The fine control needed to move the laser-spot position from the cathode to the aperture surface was accomplished with a piezo-mounted mirror in the gun periscope module. In this way, pure photoemission from either the cathode or the aperture surface could be achieved (Fig, 1d-f).[18] Based on the aperture diameter and the laser-spot size, the approximate minimum physical planar peak separation of the two photoemitting regions is ~0.6 mm. For positions between the two ideal settings, photoemission from the aperture and the cathode can be observed (Fig. 1e).[11,18]

Alignment of the Wehnelt-aperture photobeam can be done using conventional methods, with gun shift and tilt requiring the most significant changes relative to on-axis emission; in addition to the images shown here, acquisition of pulsed-beam diffraction patterns and generation of nanoscale probe beams illustrate the viability of this configuration.[18]

We next conducted a series of pump-probe UEM imaging experiments on the 2$H$-MoS$_2$ specimen (Fig. 2a; Multimedia available online). We selected a region of interest (ROI) from which CAP dynamics were probed with each photoemitting geometry. (When comparing dynamics within ROIs captured using the different geometries, ROI position and size were matched across the different image series using drift correction and cross-correlation methods.) Again, while the *SNR* for the Ni aperture was lower than that for the Ta cathode, the contrast patterns were nevertheless usable and comparable (Fig. 2b,c). Matched sub-ROIs (*e.g.*, the red rectangles in Fig. 2b,c) were used to generate space-time contour plots (STCPs) of the CAP dynamics so that relative onset times and electron-packet TOF could be determined.[34,35]

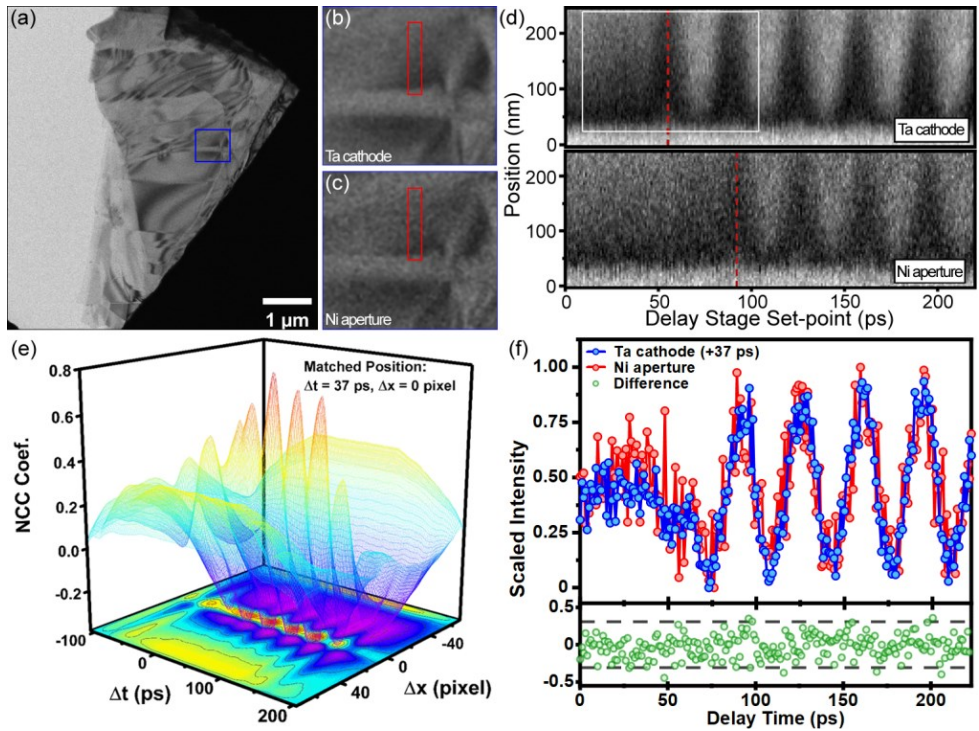

**Figure 2.** (Multimedia available online) *r*-TOF of photoelectron packets from the Ta cathode and the Ni Wehnelt aperture determined with CAP dynamics. Each UEM image series captures dynamics at 2-ps steps spanning 212 ps in total slowed by a factor of $1.75 \times 10^{10}$. (a) Bright-field image of the 2*H*-MoS$_2$ flake. The ROI within which CAP dynamics were quantified is outlined in blue. (b,c) Magnified ROIs of UEM images generated using the Ta cathode and the Ni aperture, respectively. STCPs were generated from the rectangular regions outlined in red. (d) STCPs from scans generated using the Ta cathode and the Ni aperture (upper and lower panel, respectively). Peaks of the first period of image-contrast oscillation are marked with vertical red dashed lines. The region used as a template for the cross-correlation calculations is outlined in white. (e) Normalized cross-correlation (NCC) coefficient between the Ta cathode STCP template and the Ni aperture STCP. The maximum coefficient of 0.77 occurs at $\Delta t$ = 37 ps and $\Delta x$ = 0 nm. (f) Comparison of the time-dependent scaled image intensities generated with the two sources. Intensities were tracked at the same relative pixel positions within the sub-ROIs. Data generated

with the Ta cathode are shifted +37 ps to illustrate the maximum correlation. Statistical noise level (horizontal dashed lines) is compared to the image-intensity difference between the two sets.

The STCPs generated from the identical sub-ROIs in Figure 2b,c both display a coherent contrast oscillation over the selected delay-stage temporal window (0 to 212 ps, sampled every 2 ps; Fig. 2d). This behavior arises from photoexcitation of CAPs that then propagate along the *c*-axis layer-stacking direction in the flake.[30,35–43] By identifying and comparing a shared feature of the STCPs (*e.g.*, the peak position of the same period of contrast oscillation – see the vertical red-dashed lines in Fig. 2d), a temporal shift in the onset of CAP dynamics imaged with each source becomes apparent.[35] See Figure-2 Video 1 for a side-by-side comparison of representative UEM image scans generated with the two photoemitting geometries.

To quantify the shift seen in Figure 2d, we selected a distinctive region of the Ta-cathode STCP as a template. Using this template, we performed a normalized cross-correlation (NCC) with the Ni aperture STCP.[44] We selected a template region that contains both pre- and post-excitation responses owing to the distinct asymmetry in the STCP pattern. The NCC coefficient, *γ*, is defined in Equation 4.

$$\gamma(u,v) = \frac{\sum_{x,y}[f(x,y)-\bar{f}_{u,v}][t(x-u,y-v)-\bar{t}]}{\left\{\sum_{x,y}[f(x,y)-\bar{f}_{u,v}]^2 \sum_{x,y}[t(x-u,y-v)-\bar{t}]^2\right\}^{1/2}} \qquad (4)$$

Here, *f* refers to the image, while *t* refers to the template with coordinates (*u,v*). All summations are over *x,y* under the window containing the template. The NCC coefficient serves as a measure of the similarity between the image (*f*) and the template (*t*), with a coefficient of one indicating a perfect correlation (*i.e.*, *f* is identical to *t*). Accordingly, the NCC coefficient generated by comparing the STCPs is a measure of the *spatiotemporal* similarity of the CAP dynamics probed with both photoemission geometries. Applying this to the STCPs in Figure 2d, a maximum NCC coefficient of 0.77 occurs for a temporal shift of the Ta-cathode STCP by +37 ps and a 0-nm spatial shift (Fig. 2e). (The 0-nm spatial shift is indicative of the same CAP dynamics occurring within identical sub-ROI sizes and locations.) This offset agrees with the observed delay stage set-point positions of the peaks of the first period of oscillation (55 ps for the Ta cathode vs. 92 ps for the Ni aperture). It is noteworthy that the offset seen here is comparable to the 17-ps offset observed for shank vs. tip apex photoemission from a tapered 16-μm flat $LaB_6$ cathode for a 100-V Wehnelt bias.[11]

We also compared the scaled intensities of identical points within the sub-ROIs. By shifting the Ta-cathode STCP by +37 ps, the intensity differences between the two points as a function of time are at a minimum (Fig. 2f), in agreement with the amount of shift found using the NCC method. Indeed, the minimized difference falls almost completely within the calculated image noise level across the entire 212-ps range (Fig.-2f lower panel). Following the approach taken for Equations 1-3, we approximated the noise level at pixel position $i,j$ in the image ($I$) as $\sigma[I(i,j)] \cong \frac{I(i,j)}{SNR}$. The statistical noise level was determined by averaging the absolute intensity at the point of interest in the sub-ROIs from each image acquired up to 20 ps before the onset of dynamics. The resulting values were used to calculate $\sigma[I(i,j)]$. Because data generated with the different sources have different image counts, associated *SNRs* were used. Each value was normalized by the maximum intensity value in the corresponding $I(t)$ data sets.

To determine if the observed temporal shift was photoemitting-geometry dependent rather than specimen-ROI dependent, we analyzed an additional five ROIs. The results from three of the five are shown in Figure 3. (Results from all five were consistent – we chose to display three simply for clarity and conciseness.) A specimen region different from that in Figure 2 was selected (Fig. 3a). The three ROIs within this region selected for display are indicated in Figure 3b. ROIs amenable to analysis were selected based on the presence of clear CAP dynamics in the image series, which manifested as discernable bend-contour oscillations. As can be seen in the Figure-2 Video, the sensitive nature of bend-contour position to changes in relative orientation of incident electron wave vector and crystal lattice can produce rich dynamics across a large swath of the specimen.[34,45–48]

As can be seen in Figure 3c, CAP dynamics over the delay stage set-point range of 212 ps are shifted for all ROIs when comparing the Ta cathode and the Ni aperture geometries. The shift is the same sign and magnitude for all ROIs, and it is the same as that seen for the ROI shown in Figure 2. Statistical analysis of all six ROIs probed in this study returns a $\Delta t$ shift of $37.3 \pm 0.5$ ps, with the onset of dynamics probed with the Ni aperture being shifted to a later delay-stage set-point position. For our pump-probe configuration, we scan the pump beam in time relative to a fixed probe beam. Here, we set the longest pathlength of the pump beam along the delay stage as $t = 0$ ps. Positive delay-stage set-points correspond to shortening pathlength and relatively earlier arrival times at the specimen plane, as occurs for electron packets photoemitted from the Wehnelt aperture.

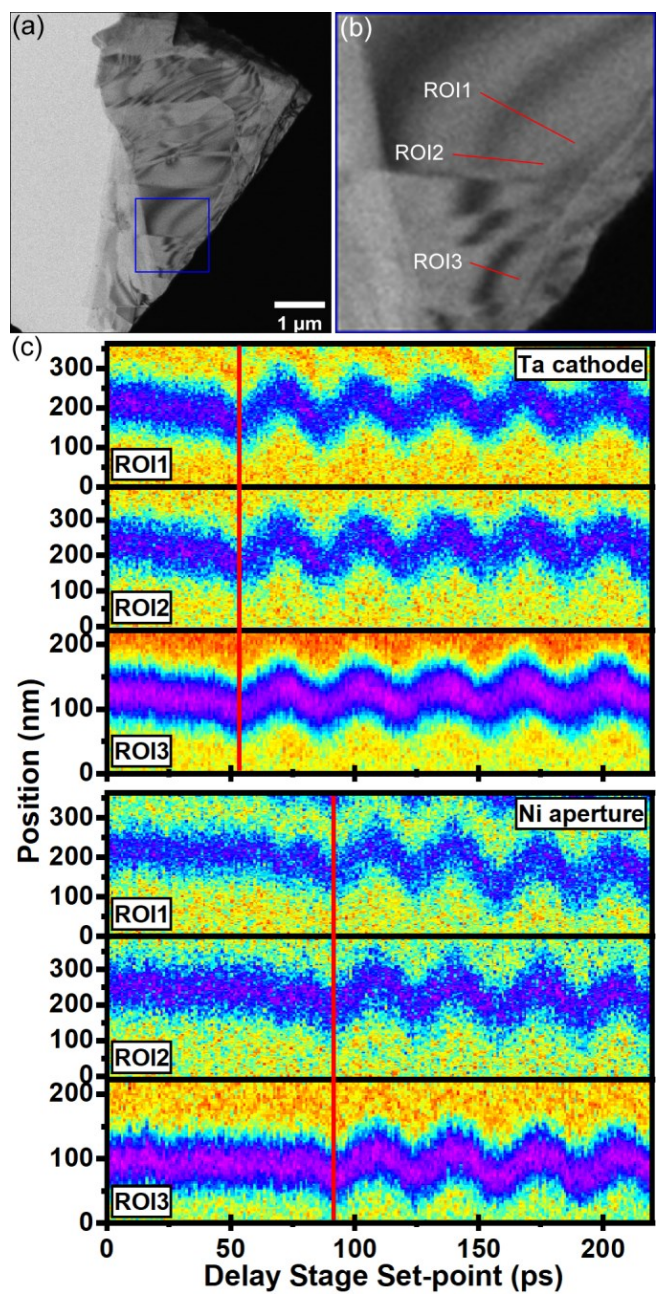

**Figure 3.** *r*-TOF of photoelectron packets emitted from the Ta cathode and Ni Wehnelt aperture, as determined with CAP dynamics. (a) UEM image of the flake with the overall region outlined in blue. (b) Magnified view of the overall region of interest showing the three specific ROIs from which CAP dynamics were quantified. (c) STCPs from the three specific ROIs generated with the Ta cathode (upper three panels) and the Wehnelt aperture (lower three panels). Onset of dynamics for each series is marked with a red vertical line.

To better understand the origins of the temporal shift, we performed particle-tracing simulations with GPT.[29] Electrostatic field maps were generated with Poisson Superfish for the Tecnai Femto gun architecture and photoemission geometries studied here.[28] Trajectories of $5 \times 10^4$ non-interacting electrons emitted from both sources were tracked. The dimensions of the emitters were the same as those used in the experiments. For electrons passing through the limiting X-ray aperture, the arrival time at a virtual screen 0.35 m from the Ta cathode surface (approximate entrance to the illumination system) was defined as the electron TOF. Counts at the virtual screen were used to define the collection efficiency (CE). Initial kinetic energy spreads were calculated using Fermi Dirac statistics, initial momenta were assigned according to an azimuthally integrated $\cos\theta$ distribution, and initial spatial coordinates were Gaussian. See Refs. 16 and 17 for details.

The simulation results are summarized in Figure 4. There are several features worth noting. Most importantly, envelopes of the TOF distributions show that electrons emitted from the Ta-cathode position do indeed arrive at the virtual screen later than those emitted from the aperture, in agreement with experiments. For a set-back distance of 0.35 mm, peak positions of the TOF distributions differ by 25.3 ps, which is 12 ps less than the experiments. This difference has several likely origins. First, no electron-electron interactions were included in the simulations. This was done so that baseline behavior could be established prior to increasing the complexity. In the experiments, hundreds of electrons populated each individual packet. Thus, repulsion could impact momenta such that larger dwell times occur in the gun. Second, simulations were carried out only to the entrance of the illumination system, while CAP dynamics reflect behaviors indicative of accumulated interactions from the source to the specimen. Additional dispersive surfaces encountered by packets could further exacerbate the temporal offset.

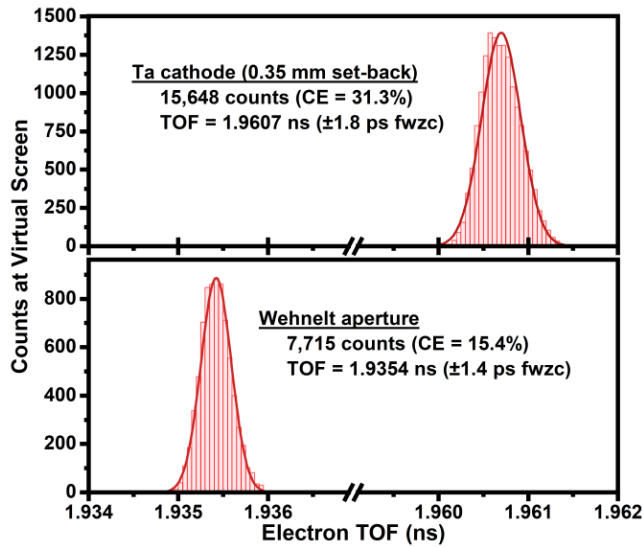

**Figure 4.** Simulated electron TOF for the Ta cathode (top) and the Wehnelt aperture (bottom) photoemitting geometries. Counts are the total number of electrons reaching the virtual screen, and CE is the percentage of the total collected divided by the total emitted ($5 \times 10^4$). The TOF is the peak position of the best-fit Gaussian to the histograms, and the range is reported as full-width at zero-counts (fwzc).

Third, deviation from expected material work function could impact the initial kinetic energy spread in ways not captured here. Indeed, the impact of initial kinetic energy, combined with photoemitting geometry, is reflected in the observed difference in simulated CE and the TOF full-width at zero-counts (fwzc). For example, for the conditions used here, the CEs and TOF fwzc were 31.3% vs. 15.4% and 1.8 ps vs. 1.4 ps for the Ta cathode and Wehnelt aperture, respectively. These differences are mainly due to the off-axis geometry acting as an unintended energy filter – electrons emitted with relatively high transverse momenta are less likely to pass through the limiting X-ray.[16,17] Thus, this will lead to a reduced CE and a *narrowed* temporal distribution for the off-axis geometry, as seen in the simulation results in Figure 4. These and other effects will be the subject of future systematic studies of this unconventional photoemitting geometry.

In conclusion, we have shown that a clear, quantifiable connection exists between the photoemitting position in the electron-gun region and the onset of structural dynamics in fs laser-driven UEM. We accomplished this by using *in-situ* photoexcited coherent phonon dynamics and an unconventional photoemitting geometry recently shown to be a viable means of stable and robust pulsed electron-beam generation.[18] Supporting particle-tracing simulations not only agreed with the experiments but also uncovered additional elements worthy of further investigation. The results reported here are expected to impact current efforts dedicated to defining the resolution limits of TEG-based UEMs and to pushing these limits to the combined Å-fs range for probing the widest-possible range of materials responses at the fundamental spatiotemporal limits of structural dynamics.[5,6]

**Acknowledgments**

This material is based upon work supported by the U.S. Department of Energy, Office of Science, Office of Basic Energy Sciences under Award No. DE-SC0018204. This work was supported partially by the National Science Foundation through the University of Minnesota MRSEC under Award Number DMR-2011401.

**Author Contributions**

**Jialiang Chen:** Data Curation; Formal Analysis; Investigation; Methodology; Validation; Visualization; Writing – original draft. **Simon Willis:** Investigation; Methodology. **David Flannigan:** Conceptualization; Funding Acquisition; Methodology; Project Administration; Resources; Supervision; Visualization; Writing – review & editing.

**Author Declarations**

The authors have no competing interests to declare.

**Data Availability**

The data are available from the corresponding author upon reasonable request and will also be deposited in the Data Repository for U of M (DRUM).